\documentclass[usegraphicx]{mn2e}
\usepackage{amssymb, amsmath}
\usepackage{graphicx}

\def \bv {{\bf v}}
\def \bu {{\bf u}}
\def \br {{\bf r}}

\def \bnabla {{\boldsymbol{\nabla}}}
\def \bnablaperp {{\bnabla_{\negthickspace\perp}}}

\begin{document}
\title[Kinematic Bias in Cosmological Distances]{Kinematic Bias in Cosmological Distance Measurement}
\author[Kaiser \& Hudson]{Nick Kaiser \& Michael J. Hudson \\
$^1$Institute for Astronomy, University of Hawaii \\
$^2$Department of Physics and Astronomy, University of Waterloo}
\maketitle

\begin{abstract}
Recent calculations using non-linear relativistic cosmological perturbation theory
show biases in the mean luminosity distance and distance modulus at low redshift.  
We show that these effects may be understood very simply 
as a non-relativistic, and purely kinematic, Malmquist-like bias, and we describe how the
effect changes if one averages over sources that are limited by apparent magnitude. 
This effect is essentially identical to the distance bias from small-scale random velocities
that has previously been considered by astronomers, though we find that
the standard formula overestimates the homogeneous bias by a factor 2.
\end{abstract}


\section{Introduction}

It is well known that the local rate of expansion $H_0$ is significantly perturbed,
at linear order, by peculiar velocities associated with the growth of density perturbations.
The impact of this on cosmological parameter estimation
is quantified theoretically by calculating the covariance of the 1st order
velocity field which is given in terms of the power spectrum of density fluctuations 
(Hui \& Greene 2006; Cooray \& Caldwell 2006; Davis et al.\ 2011; Kaiser \& Hudson 2014).

The subject of this paper, in contrast, is the systematic {\em bias\/} in distances, and
therefore $H_0$, caused by velocities, and which is a second order effect.
This has been studied using 2nd order relativistic cosmological
perturbation theory in a number of recent papers 
(Vanderveld, Flanagan \& Wasserman, 2007;
Li \& Schwarz 2008; 
Clarkson, Ananda \& Larena 2009;
Umeh, Larena \& Clarkson 2011;
Gasperini et al.\ 2011;
Wiegand \& Schwarz 2012;
Fanizza et al.\ 2013;
Ben-Dayan et al.\ 2012a, 2012b, 2013a, 2013b, 2014).

These papers all compute the deviation of quantities such as the mean 
luminosity distance and distance modulus (log distance), averaged over
a surface of constant redshift, from that which would apply in a homogeneous universe.
Second order perturbation theory is being used in order to explore
the regime of non-linear gravitational dynamics. Most of these papers
describe the effect as backreaction from the
formation of structure, though the term may be being used in a 
relatively broad sense compared to the narrow definition
as the effect of non-commutativity of spatial averaging and time evolution
deriving from the non-linearity of Einstein's equations.

Quantitative predictions in the context of conventional structure formation
models are provided in e.g.\ figure 6 of Ben-Dayan et al.\ 2013b
which shows that the bias falls off inversely as the
square of the redshift; that the fractional perturbation to the mean distance 
$\delta_d \equiv \langle \delta d_L \rangle / d_L$ is positive,
and that the perturbation to the mean flux density $\Phi$ is negative with 
$\delta_\Phi \equiv \langle \delta \Phi \rangle / \Phi \simeq - 0.5 \delta_d$.
Further, according to Ben-Dayan et al.\ 2014 (hereafter BDMS14), 
for low redshift $z \ll 1$ the mean flux density perturbation is 
given in terms of $\langle v^2 \rangle$, the total variance of the 
first order line-of-sight peculiar velocity, by $\delta_\Phi = - \langle v^2 \rangle / c^2 z^2$,
and they give the bias in the distance modulus 
$\mu = 5 \log d_L = (5 / \ln 10) \ln d_L$ as $\langle \delta \mu \rangle = (7.5 / \ln 10) \langle v^2 \rangle / c^2 z^2$.

There are two surprising features of these results if they are assumed to be caused by
inhomogeneity affecting the evolution of the averaged universe.  First, a cosmological effect
would be expected to grow with increasing redshift rather than decrease.
Second, one would expect perturbations to distance, distance modulus and flux density to be
related by $\langle\delta \mu \rangle = (5 / \ln 10) \langle\delta d_L \rangle / d_L$ and 
$\langle \delta \Phi \rangle / \Phi = -2 \langle \delta d_L \rangle / d_L$, just 
as for an individual `standard candle'.
The relations between these quantities obtained from perturbation theory are quite different,
and suggest that the cause of these effects are fluctuations.  In that case, 
the usual relations for a standard candle would not apply, simply because of the non-commutativity of
averaging and non-linear transformations; the mean of the square of a fluctuating
quantity, for example, is of course not the same as the square of the mean.
The effect of fluctuations and the non-linearity of the relationships between $d_L$, $\mu$ and $\Phi$
was discussed by BDMS14 who noted that the bias in $H_0$ depends on the
observable used, and by Ben-Dayan et al.\ 2013a, who argued for using the flux density $\Phi$
in $H_0$ measurements, claiming this to be the least sensitive to fluctuations.

Statistical biases in distance estimation, often associated with the names
Eddington (1914) and Malmquist (1920), have been known and widely studied
for a long time, in the context of both cosmological parameter estimation
and measurements of large-scale peculiar motions or `cosmic flows'.  Substantial biases
may result from the typically $\sim 20$\% uncertainty in 
luminosity distance estimators for galaxies
such as are obtained from the Tully-Fisher (TF) relation for spirals (Tully \& Fisher 1977)
and from the `fundamental plane' (FP) for elliptical 
galaxies (Djorgovski \& Davis, 1987; Dressler et al.\ 1987).
In particular, distance estimates to galaxies may suffer so-called 
`homogeneous Malmquist bias' in that field
galaxies in some range of estimated distance will tend to have 
true distances that are, on average, systematically
enhanced as more galaxies are scattered inward from larger
distances than outward from smaller distances (see Lynden-Bell et al.\ 1988; Willick 1994; and the reviews of 
Faber et al.\ 1994 and Strauss \& Willick 1995 for more details).
Lynden-Bell et al.\ 1988 showed that with a log-normal model
for the distribution of distance errors the mean log-distance in a spatially
homogeneous universe would be biased upward by $\delta \ln d = 3 \Delta^2$
where $\Delta^2$ is the fractional distance error variance.

This particular kind of bias may be avoided by 
considering the mean peculiar {\em displacement\/} in redshift-space (where neighbouring
sources have, to a good approximation, the same distance) rather than 
the peculiar motion in estimated distance space (Schechter 1980).   This bias is also
not particularly relevant to the calculations above as they
effectively assume perfect standard candles.
What {\em is} relevant is the residual bias
that persists after the bias from distance errors has been eliminated.  This is driven
by small-scale velocity dispersion which causes a scatter in
the true distance for objects at the same redshift.  
This was first considered by Lynden-Bell (1992) who calculated the shift in the
mode of the distribution of log-distances for objects of a given recession velocity.  Specialising to
uniform density and ignoring selection effects and streaming motions gives
$\delta \ln d = 3  \sigma_v^2 / c^2 z^2$ where $\sigma_v$ is the velocity dispersion.
Willick et al.\ 1997 also found, under the same simplifying assumptions, 
that velocity dispersion induces a bias in the
apparent magnitude (or distance modulus) of sources of given redshift
of $\delta m = 3 \times (5/\ln 10) \times \sigma_v^2 / c^2 z^2$.
And both of these are just what one would expect from the Lynden-Bell et al.\ 1988
formula for the standard Malmquist effect with fractional distance
error variance $\Delta^2 = \sigma_v^2 / c^2 z^2$, which seems very reasonable.

Lynden-Bell (1992) and Willick et al.\ (1997) considered the
effect of motions on small-scales that are modelled as spatially
incoherent with galaxies behaving like a gas of particles with a Maxwellian velocity
distribution.  This is very different from the modelling of velocities in perturbation theory,
where the motion is like that of a smooth, cold fluid. 
But otherwise the results are
qualitatively the same in that the bias falls off as $1/z^2$ and is proportional to the
mean square velocity.  This might lead one to suspect that the perturbation theory
results are simply the analogue of Malmquist-like bias from small scale motions; which
are entirely a consequence of kinematics and statistics. On closer inspection,
however, there is a difference in that $\delta m$ is twice
as large as the $\delta \mu$ of Ben-Dayan et al.\ 2014.

In this paper we will explore these biases further.  The questions we address are:
To what extend can the perturbation theory results be understood in terms of kinematics and statistics?
Why does there appear to be a difference between the effects of perturbative
flows and small-scale incoherent motions?  Is this some subtle relativistic effect?  Or might
it perhaps derive from some significant difference between the statistical properties of
small-scale and large-scale motions? or from the neglect of density perturbations associated
with the latter?
Another question is why the perturbation theory analysis result for
the bias on local measurements of $H_0$ is determined by
the total velocity variance, including that from very long wavelength perturbations, 
when one would expect only relative motions -- which for super-survey scale modes
are suppressed -- to appear.

\section{Malmquist Bias from Large-Scale Coherent Flows}
\label{sec:coherentflows}

Here we will calculate the kinematic bias arising from `coherent flows'
or `streaming motions'; these being the focus of the relativistic perturbation
theory calculations.  We consider small-scale `thermal' motions later.
Since we are interested in the low redshift regime $z \ll 1$ we work in
flat, empty space and, we will also ignore special relativistic effects as
the effects of interest here are generally of order $\sim (v / cz)^2 \gg (v/c)^2$.  

We first consider the bias in the distance and related quantities
when averaged over the surface of constant redshift as this is simple, illustrates
the key features of the phenomenon, and is what was considered in the relativistic
perturbation theory studies.  We then generalise the analysis to the
more realistic case where we average these quantities over sources.

\subsection{Area Averaged Bias}
\label{subsec:area_average}

We imagine an ensemble of realisations of a smooth field of test
particles that have a spatially continuous velocity field that
consists of a Hubble flow $H \br$ plus a statistically homogeneous
random velocity perturbation field, and where one particle is selected
at random as the observer and is taken to lie at the origin of
spatial coordinate system.  Let the velocity with respect to this
observer be $\bu(\br)$ and define the peculiar velocity 
$\bv = \bu - H \br$.
Let us further assume, in the spirit of perturbation analysis,
that the amplitude and scale length for perturbations in the peculiar
velocity are such that there is a unique mapping from velocity
(or redshift) space to real space; i.e.\ all particles in some
region of redshift space have the same peculiar velocity.  Working
in units such that both the speed of light $c$ and the expansion
rate $H$ are unity, the distance is $d = |\br| = z - v$ where $v$
is the line-of-sight component of the peculiar velocity.

Consider a cone of infinitesimal solid angle $d \Omega$.  In redshift space, the
intersection of this cone and a constant-$z$ surface has area $dA_z = z^2 d \Omega$.
That two dimensional surface maps to surface element in real
space that will lie at a perturbed distance $d = z - v = z (1 - v/z)$ and which
will, in general, be slightly tilted relative to the line of
sight as there will, in general, be some gradient of $v$ transverse to
the line of sight $\bnablaperp v$.  The surface element area in real space is then
\begin{equation}
dA_r = (1 - v/z)^2  (1 + |\bnablaperp v|^2 / 2) dA_z .
\end{equation}

The average of the fractional perturbation to the distance $\delta_d = (d-z)/z = -v/z$ over
a solid angle $\Delta \Omega$, weighted by real-space area, is then
\begin{equation}
\overline{\delta_d} 
= \frac{\int d\Omega (1 - v/z)^2 (1 + |\bnablaperp v|^2 / 2)(-v/z)}
{\int d\Omega (1 - v/z)^2 (1 + |\bnablaperp v|^2 / 2)} .
\label{eq:delta_d_bar_area_average}
\end{equation}

We wish to evaluate $\overline{\delta_d}$ accurate to second order in velocities.  Since there
is a factor $v/z$ in the numerator, that means we need only keep first order
terms in the denominator, and we can completely ignore the transverse derivative terms
as they appear only at third order, to give 
\begin{equation}
\overline{\delta_d} = 
\int \frac{d\Omega}{\Delta\Omega} \left\{- \frac{v}{z} + \frac{2v^2}{z^2}
- \frac{v}{z} \int \frac{d\Omega'}{\Delta\Omega}  \frac{2 v'}{z} \right \} 
\label{eq:delta_d_bar_expansion_area_average}
\end{equation}
the last factor here allowing for correlation between the
numerator and denominator in (\ref{eq:delta_d_bar_area_average}).

The integrals here are evaluated on the surface  $z = $ constant  i.e.\ on the perturbed 
surface in real space $d = z - v$.  Working to second order precision, $\overline{\delta_d}$
is given in terms of quantities on the constant distance surface $d = z$ using 
$v(z - v) = v(d=z) - v dv/dd + \ldots = v - (1/2) d v^2 / dz + \ldots$ (we can ignore
the effect on the second order terms above as the change in these is third order in $v$).

Taking the ensemble average, which we will denote by $\langle \ldots \rangle$, 
the expectation value of the first order term here vanishes as the velocity
is equally likely to be positive as negative -- this is equally true in real-space and
redshift-space since, like $v(d)$, $d v^2 / dz$ is equally likely to be positive or negative -- with the result
\begin{equation}
\begin{split}
\langle \overline{\delta_d} \rangle & = \frac{2\langle v^2 \rangle}{z^2}  -
\frac{2}{z^2}  \int \frac{d\Omega}{\Delta\Omega} \int \frac{d\Omega'}{\Delta\Omega} \langle v v' \rangle \\
& \quad\quad = \frac{1}{z^2} \int \frac{d\Omega}{\Delta\Omega} \int \frac{d\Omega'}{\Delta\Omega} \langle (v - v')^2 \rangle .
\end{split}
\label{eq:delta_d_avg}
\end{equation}

The last expression above makes it clear that $\langle \overline{\delta_d} \rangle > 0$ so
the mean distance is biased upwards.
It also shows that, for an averaging area that
subtends a small solid angle $\Delta \Omega \ll 1$, only
velocities caused by density perturbations with scale comparable
to or smaller than the averaging region contribute significantly to the bias; for perturbations
much larger than the averaging region size the velocity will vary little within
the area so $v' \simeq v$ and the bias is strongly suppressed.

If instead of the perturbation to the distance, which is linear 
in $v$ (for given $z$), we calculate the perturbation to some 
observable $X$ that is a non-linear function of distance like the flux-density
or the distance modulus then we need to include the second order term in the expansion of $X$
expressed as a function of $v/z$.  If the perturbation is 
$\delta X = a v/z + b v^2 / z^2 + \ldots$ then
we simply replace the factor $(-v/z)$ in (\ref{eq:delta_d_bar_area_average}) 
by $a v/z + b v^2 / z^2$ and performing the same expansion -- dropping
terms that are cubic or higher in the velocity -- and ensemble 
averaging that led to (\ref{eq:delta_d_bar_expansion_area_average}) 
and then to (\ref{eq:delta_d_avg}) now gives \begin{equation}
\langle \overline{\delta X}  \rangle = (-2 a + b) \frac{\langle v^2 \rangle}{z^2} 
+ \frac{2 a}{z^2}   \int \frac{d\Omega}{\Delta\Omega} 
\int \frac{d\Omega'}{\Delta\Omega} \langle v v' \rangle .
\label{eq:delta_X_avg}
\end{equation}

We can use this to give the fractional perturbation to the flux density of standard sources.
These have $\Phi(d) \propto 1 / d^2$ so $\Phi(d) = \Phi(z) (1 - v/z)^{-2}$
and $\delta_\Phi \equiv (\Phi(d) - \Phi(z)) / \Phi(z) = (1 - v/z)^{-2} - 1 = 2 v/z + 3 v^2 / z^2 + \ldots$
so the ensemble average of the area averaged flux density perturbation $\delta_\Phi$ is given by
(\ref{eq:delta_X_avg}) with $a = 2$, $b = 3$ or
\begin{equation}
\langle \overline{\delta_\Phi} \rangle = - \frac{\langle v^2 \rangle}{z^2} +
\frac{4}{z^2}   \int \frac{d\Omega}{\Delta\Omega} \int \frac{d\Omega'}{\Delta\Omega} \langle v v' \rangle .
\label{eq:delta_Phi_avg}
\end{equation}

Similarly, the perturbation to the distance modulus (DM) $\mu \equiv 5 \log_{10} d$ is 
$\delta\mu = \alpha \ln (1 - v/z) = - \alpha (v/z + v^2 / 2 z^2 + \ldots)$
with $\alpha \equiv 5 / \ln 10 \simeq 2.17$, so the ensemble average of the area average of
$\delta \mu$ is given by (\ref{eq:delta_X_avg}) with $a = -\alpha$, $b = - \alpha / 2$ or
\begin{equation}
\langle \overline{\delta\mu} \rangle = \alpha
\left[ \frac{3 \langle v^2 \rangle}{2z^2} -
\frac{2}{z^2}   \int \frac{d\Omega}{\Delta\Omega} \int \frac{d\Omega'}{\Delta\Omega} \langle v v' \rangle \right] .
\label{eq:delta_mu_avg}
\end{equation}

Note that in both of these cases, in contrast to (\ref{eq:delta_d_avg}), there is not complete suppression of the 
effect of perturbations on scales larger than the averaging area.

If we take the averaging area to cover the entire sky, and assume that the redshift is sufficiently
large that the distance to this shell is much greater than the coherence scale for the velocity
fluctuations then the second term involving $\langle v v' \rangle$ in each of equations 
\ref{eq:delta_d_avg}, \ref{eq:delta_Phi_avg} \& \ref{eq:delta_mu_avg}
will be much smaller than the first term and we have
\begin{equation}
\begin{split}
\langle \overline{\delta_d} \rangle & = 2 \langle v^2 \rangle / z^2 \\ 
\langle \overline{\delta_\Phi} \rangle & = - \langle v^2 \rangle / z^2 \\
\langle \overline{\delta\mu} \rangle & = (7.5 / \ln(10)) \langle v^2 \rangle / z^2 . \\
\end{split}
\label{eq:delta_bar_collection}
\end{equation}
These are identical to the low-$z$ limit expressions of BDMS14.
So the relatively large low-$z$ effects are not in an
essential way a result of non-linearity of gravitational dynamics (relativistic
or Newtonian) as they are fully accounted for by kinematics and statistics. 
We believe, of course, that the velocities we observe are
really caused by gravity, and non-linear structure is involved, 
but our point here is that the same bias would be
found if one were observing test particles of negligible mass
with peculiar motions caused by non-gravitational forces.

\begin{figure}
\begin{center}
\includegraphics[width=85mm]{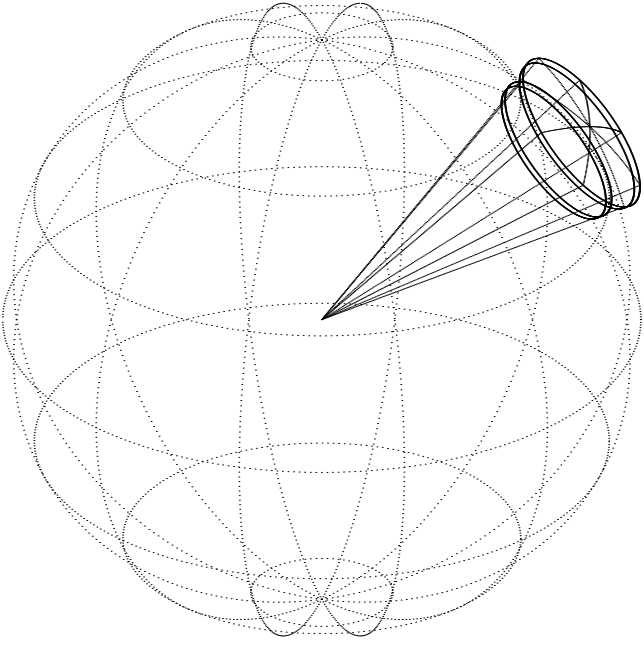}
\end{center}
\caption{Dotted lines are lines of longitude and latitude on the
surface of constant redshift.  On this surface,
peculiar velocities are equally likely to be positive as negative.  The cone
illustrates how a section of this sphere maps to real space for the case of a negative
peculiar velocity.  The section is pushed out radially away from the observer
-- who resides at the centre of the sphere -- and consequently is expanded
in area.  Similarly, for a positive peculiar velocity the section
would be compressed.  The result of this is that the average of the distance,
when weighted by real-space area, is positive.  This is the cause of the
bias found in the relativistic perturbation theory analyses.  
More relevant to real observations is the bias in distance averaged
over the sources that lie in a shell of given redshift.  We consider
this in \S\ref{subsec:galaxy_average}. There we find that there are some
relatively minor differences that arise from the clustering of sources
and from the Jacobian involved in transforming volumes from redshift to
real space, but the main difference is that the generalisations
of (\ref{eq:delta_bar_collection}) have different numerical
pre-factors when the sources are subject to selection based on flux density.}
\label{fig:spherefig}
\end{figure}

These Malmquist-like biases are easy to understand.  The perturbation to the
mean distance, for example, comes about because even though the velocity
field on a sphere of constant-$z$ is equally likely to be 
positive or negative, so as many areas (or solid angle elements at the observer) 
get pushed out as get pushed in in distance-space, those that get pushed
out to larger $d$ get pushed in the radial direction and so get
expanded in area by a factor $(1 - v/z)^2 \simeq 1 - 2 v/z$ (see figure \ref{fig:spherefig}).
Similarly those that get displaced inwards get compressed.  The result is a rectification
of the real-space area averaged distance.  The different numerical factors
for the other variables comes about simply because they are non-linear functions
of the distance.

BDMS14 noted that the above imply that the bias in $H_0$ 
obtained from the area-averaged flux density is a factor 3 lower than that obtained
from averaging the distance modulus.  The above analysis shows that one
can do even better by averaging $\Phi^{3/2} \propto (1 - v/z)^{-3}$ since
this gives $a = 3$ and $b = 6$ so $-2a + b = 0$ and,
in the approximation that the depth is greater than the coherence scale
used to obtain (\ref{eq:delta_bar_collection}), the bias vanishes.

We would emphasise that, according to our analysis, the simple results (\ref{eq:delta_bar_collection})
are only valid for velocity perturbations with coherence scale less than the distance.  
But at the same time the effects are really only significant at low redshift
because of the $1/z^2$ scaling.  
For realistic power spectra there is significant contribution to the 
velocity variance from quite large scales; certainly extending to tens if not hundreds of Mpc,
so except for observations at much greater distance -- where the effects
rapidly become uninterestingly small -- one should not use these formulae 
with the total velocity variance computed in the usual way from the matter power spectrum
rather one should use equations \ref{eq:delta_d_avg}, \ref{eq:delta_Phi_avg} \& \ref{eq:delta_mu_avg}
that incorporate the terms involving the velocity correlation function $\langle vv' \rangle$.

It is also important to realise that we have defined the peculiar velocity here such that the
velocity of the observer vanishes.  Thus the variances and co-variances in these
equations are of velocities relative to the observer, which in practice is usually 
taken to mean relative to the velocity of the local group (LG), since it is the LG peculiar velocity, unlike the motion of the earth or the sun, that is thought to best reflect the gravitational acceleration from large scale structures.  This eliminates the effect of
perturbations on scales much greater than the survey depth which would otherwise give
unphysical effects if the total velocity variance were used.  

This is somewhat at odds with BDMS14, and deserves some clarification.
Their equations 5,6 give a bias that depends on the total velocity
dispersion, including a contribution that comes from modes which are
larger than the survey scale spanned by the target objects (in their
case $H_{0}$ calibrators). This is the dispersion of one component of
the velocity of a galaxy relative to the `cosmic-frame', as is thought
to be well approximated by the frame in which the CMB has zero dipole
(since any intrinsic dipole is usually thought to be very small). In
their discussion of this BDMS14 say that they remove the motion of the
observer since the observations are usually quoted in the CMB frame,
corresponding to $\bv_0 = 0$, and that a non-vanishing observer
velocity would nearly double the effect.  This doubling seems to us to
be misleading.  The observer velocity is not zero in the CMB frame --
the LG is moving at about 600 km/s in that frame -- but the CMB frame
is not of much relevance here as the results should be independent of
any frame that the observers choose to refer the observations to.  Our
formulae, including the correlation function $\langle vv' \rangle$,
refer to ensemble averages and, if one had no idea how the LG motion
originated, then these should be in the LG frame.  Working in the LG
frame would indeed increase the co-variance from perturbations on
scales smaller than the survey scale, though the effect of motions on
larger scales would still be suppressed.

But there is a difference between the variance of the motions of
different source regions and our motion, which has a variance in an
ensemble sense, but we only sample one realisation of the ensemble
(though it is a realisation of all three components of the velocity,
not just one). The exact impact of the LG's motion depends on the
depth of the gravitational sources that are responsible for its
motion: if these sources are deeper then the $H_{0}$ secondary
calibrators themselves, then the $H_{0}$ calibrators and the LG motion
share the same bulk velocity and so, by operating in the LG frame,
these super-survey modes disappear, as noted above.  If on the other
hand, the source of the LG's motion is very local to the LG itself
(for example, a very nearby attractor such as Virgo), then, when
operating in the LG frame, the LG motion induces a coherent dipole
pattern (see Kaiser \& Hudson 2014 and references therein).  This
coherent dipole is different in character to the less-coherent
distortion due to the motions of the $H_{0}$ calibrators.

In practice, however, the LG's motion arises from gravitational
sources over a wide range of distances, so the true situation is more
complicated than the two scenarios sketched above.  Fortunately, by
mapping out the distribution of nearby galaxies with an all-sky
redshift survey and predicting peculiar velocities via linear
perturbation theory, we now have a good idea of the gravitational sources
responsible for much of the LG's motion (e.g.\ Erdogdu et al.\ 2006;
Lavaux \& Hudson 2011, Carrick et al 2014).  Consequently, because in
practice these surveyed volumes contain within them the secondary
calibrators with which one is attempting to measure the local $H_0$,
the bias in the local value of $H_0$ could be reduced by working in
the frame of the redshift survey itself.  In other words, the solution
is to use the predicted peculiar velicities to correct for the
redshifts of the calibrators (Neill et al. 2007, Riess et al. 2011),
leaving only 150-200 km/s of peculiar velocity not well described by
linear theory (Carrick et al. 2014).
%
%

\subsection{Galaxy Averaged Bias}
\label{subsec:galaxy_average}

We now explore how the bias changes if, as is the case in reality, we average distances over
galaxies (or supernovae), rather than perform an area weighted average on 
the surface of constant redshift, and allow for the fact that such sources
are subject to selection bias.
This involves weighting by volume elements of a shell
that maps to a shell of constant thickness in redshift space, rather than by area on the surface of constant $z$, and this introduces
a factor which is the Jacobian of the real- to redshift-space transformation.
And there are additional weighting factors coming from the varying real-space density of
galaxies arising from structure and from the distant dependent selection function.

Consider a segment of a spherical shell in redshift space at 
redshift $z$ and thickness $dz$ that subtends a solid angle
$d\Omega$ at the observer, and which therefore has volume $dV_z = d\Omega z^2 dz$.
This maps to a volume in real-space 
\begin{equation}
dV_r = (1 - v/z)^2  (1 - dv/dz) dV_z
\end{equation}
where we see the Jacobian $1 - dv/dz$.  Unlike the tilt factor $1 + |\bnablaperp v|^2 / 2$,
which was ignorable, this has a first order component.

The expected number of detected galaxies is proportional to the product 
$dV_r (1 + \delta) \phi(d)$ where $\delta$ is the real-space galaxy density contrast
and $\phi$ is the selection function, which we can take to be a function of real distance
$d$, since e.g.\ effects from aberration caused by our motion changing area of
galaxies is an order $v/c$ effect and is relatively negligible. 

For $v \ll z$ we can make a first order
expansion and write the latter as $\phi(d) = \phi(z) (1 - (v/z) d \ln \phi / d \ln z + \ldots)$.
As before we shall only need to keep the first order term in the expansion of $(1 - v/z)^2$, so
we can use $(1 - v/z)^2 \phi(d) = \phi(z) (1 - (2 + \gamma) v/z + \ldots)$ 
where $\gamma \equiv d \ln \phi / d \ln z$.

The average of the fractional perturbation to the distance over
a solid angle $\Delta \Omega$, weighted by galaxy number, is then
\begin{equation}
\overline{\delta_d} 
= \frac{\int d\Omega (1 - (2 + \gamma) v/z)  (1 - dv/dz) (1 + \delta) (-v/z)}
{\int d\Omega (1 - (2 + \gamma) v/z)  (1 - dv/dz) (1 + \delta)} .
\label{eq:delta_d_bar}
\end{equation}

As before, we wish to evaluate $\overline{\delta_d}$ accurate to 
second order in perturbed quantities (now
including $\delta$ as well as velocity).  Expanding
and neglecting terms that are cubic or higher yields
\begin{equation}
\begin{split}
\overline{\delta_d} & = - \int \frac{d\Omega}{\Delta\Omega} \frac{v}{z} \left\{
1 - \frac{(2 + \gamma) v}{z} - \frac{dv}{dz} + \delta \right. \\
& \quad\quad\quad 
\left. + \int \frac{d\Omega'}{\Delta\Omega} 
\left(\frac{(2 + \gamma) v'}{z} + \frac{dv'}{dz} - \delta'\right) \right\} .
\end{split}
\label{eq:delta_d_bar_expansion}
\end{equation}

Again, when we take the ensemble average 
we will assume that the first order terms vanish by symmetry.
As already noted the product of $v$ and $dv/dz$ should average to
zero, as does the product of $v$ and $\delta$.  

But we have extra 2nd order term involving the product of
$v/z$ with $\delta'$ and $dv'/dz$.  
For a statistically homogeneous random field the expectation of
the field and its derivative at two different locations does not, in
general, vanish, nor is $\langle v \delta' \rangle = 0$ in general.
Generalising to an observable $X$ whose perturbation has the expansion
$\delta X = a v/z + b v^2 / z^2 + \ldots$ as before, the analogue of 
(\ref{eq:delta_X_avg}) is
\begin{equation}
\begin{split}
\langle \overline{\delta X}  \rangle & = (- (2 + \gamma) a + b) \frac{\langle v^2 \rangle}{z^2} 
+ \frac{a}{z^2} \int \frac{d\Omega}{\Delta\Omega} \int \frac{d\Omega'}{\Delta\Omega} \biggl\{ \\
& \quad\quad
(2 + \gamma) \langle v v' \rangle + z\langle dv'/dz - \delta' \rangle \biggr\} .
\end{split}
\label{eq:delta_X_galaxy_avg}
\end{equation}

On dimensional grounds, one might expect these new terms appearing in the double integral to
have a large contribution (as compared to the term involving $v v'$) from perturbations
with wavelength $\lambda \ll z$ since both $\delta'$ and $dv'/dz \sim v / \lambda$.
But that is misleading for the following reason.  That part of the velocity
field which derives from waves in the Fourier spectrum with wave-number $k = 2 \pi / \lambda$
has a coherence scale of order $\lambda$.  So pairs of points that have significant
correlation are restricted to have separation $\sim \lambda$, and if $\lambda \ll z$
these pairs have a separation whose direction is nearly perpendicular to the
line-of-sight.  This actually suppresses the contribution to 
$\langle \overline{\delta_d}\rangle $ from
the $v dv'/dz$ term to be smaller than that from the $v v'$ term.  
The same is true for the term involving $\langle v \delta' \rangle$.  Thus the
differences introduced by averaging over galaxies, as opposed to the simpler
averaging over areas, are small.

As was the case of averaging weighting by area, if we average over the entire 
sky and assume that this covers many `coherence-areas',
then we can ignore the double integral in (\ref{eq:delta_X_galaxy_avg})
and we have, in analogy with (\ref{eq:delta_bar_collection}),
\begin{equation}
\begin{split}
\langle \overline{\delta_d} \rangle & = (2 + \gamma) \langle v^2 \rangle / z^2 \\ 
\langle \overline{\delta_\Phi} \rangle & = - (1 + 2 \gamma) \langle v^2 \rangle / z^2 \\
\langle \overline{\delta\mu} \rangle & = (5 / \ln(10)) (3/2 + \gamma) 
\langle v^2 \rangle / z^2 . \\
\end{split}
\label{eq:delta_bar_collection_galaxy_avg}
\end{equation}

For distances of practical interest, the actual
bias involves the additional terms in (\ref{eq:delta_X_galaxy_avg}).
But the simpler expressions above are potentially useful in a situation
where large-scale motions have been modelled and corrected for, as they would
then describe any residual bias caused by un-modelled motions on smaller scales.

At any redshift the variable $d_L^n$ is unbiased for $n = -3  - 2 \gamma(z)$.
In terms of flux density $\Phi$ (and selection function $\phi$) this 
is $\Phi^{3/2 + d \ln \phi / d \ln z}$.
At the distance at which the number of galaxies per logarithmic interval of distance
is maximised -- the distance where most of the galaxies reside, in some sense --
the selection function is falling as $\phi \propto d^{-3}$ and so the
unbiased variable is $d_L^3 \propto \Phi^{-3/2}$ (as compared to the
$\Phi^{+3/2}$ that applies if there is no distance dependence selection).

\section{Malmquist Bias from Incoherent Small-Scale Motions}
\label{sec:incoherentflows}

The foregoing analysis was somewhat restricted in that it was assumed that
at each point in real-space there is a single velocity -- i.e.\ that the galaxies
move like a fluid, thus ruling out application to bound virialised
systems where there are multiple streams -- and yet more restrictive
in that it was assumed that there was a unique velocity at each 
point in redshift space; which rules out e.g.\ `triple valued' regions
in redshift space that exist around clusters.  These assumptions are reasonable only
for large scale motions.

At the other extreme, a useful and commonly used model for small
scale motions within bound structures is that these motions
are spatially incoherent with peculiar velocities drawn from a distribution function $P_v(v)dv$.
As mentioned in the Introduction, the bias caused by small-scale
motions with an assumed Maxwellian distribution (for which
the distribution of the line-of-sight velocity is Gaussian)
was considered by Lynden-Bell (1992) and by Willick et al.\ 1997,
both of whom found an effect qualitatively similar to, but twice as large as, the
bias obtained from perturbation analysis (for motions with coherence scale
less than the size of the averaging region).  This is puzzling.
Why would the result care about whether the coherence scale
is just much smaller than the averaging cell size or microscopically
small?

We now show, at least in the limit that $v \ll z$, that the
result for $\langle \overline{\delta\mu} \rangle$ in (\ref{eq:delta_bar_collection}) 
applies also to small-scale incoherent motions, and that the
use of the standard formula for the bias with distance errors replaced
by velocity errors, while entirely plausible, actually over-predicts the
effect (by a factor 2 in the case that selection is ignorable).

At any $z$, $P(d|z) \propto P(d,z) = P(z | d) P(d)$.
But $z = d + v$, so $P(z | d) = P_v(z-d | d)$.  If we assume that
the distribution of peculiar velocities $P_v$ is position
independent, then  $P(d|z) \propto P_d(d) P_v(z-d)$, from which we can compute expectation
values for distance, distance modulus etc..

With $\delta \mu = \alpha \ln (d/z)$  and assuming galaxies are uniformly distributed
in angle, but subject to some smoothly varying selection function $\phi(d)$,
so $P_d(d) = P_d(z) (1 - (2 + \gamma) v / z + \ldots)$, the mean DM for galaxies at redshift $z$ is 
\begin{equation}
\langle \delta \mu | z \rangle =
\frac{\int dv \left(1 - (2 + \gamma) \frac{v}{z}\right) P_v(v) \alpha 
\left(-\frac{v}{z} - \frac{v^2}{2 z^2} + \ldots\right)}
{\int dv \left(1 - (2 + \gamma) \frac{v}{z}\right) P_v(v)} 
\label{eq:delta_DM_int}
\end{equation}
or, keeping only terms up to second order in velocity in the numerator and only the leading
order term in the denominator,
\begin{equation}
\langle \delta \mu | z \rangle = \alpha (-\langle v\rangle / z + 
(\gamma + 3 / 2) \langle v^2\rangle / z^2) .
\label{eq:delta_DM}
\end{equation}

For the assumed centred Gaussian distribution, 
$\langle v\rangle = 0$ and (\ref{eq:delta_DM}) agrees with 
the the third of (\ref{eq:delta_bar_collection_galaxy_avg}) and, ignoring selection (i.e.\ setting $\gamma = 0)$, we have
\begin{equation}
\langle \delta \mu | z \rangle = (7.5 / \ln(10)) \langle v^2 \rangle / z^2
\label{eq:delta_mu_incoherent}
\end{equation}
in accord with the third of (\ref{eq:delta_bar_collection}) but in conflict with equation 15 of Willick et al.\ 1997 
and at odds both with equation 9.17 of Lynden-Bell (1992)
and with the seemingly reasonable analogy with Lynden-Bell et al.\ 1988,
all of which would suggest that for a uniform spatial
distribution of galaxies $\delta \ln d = 3 \langle v^2 \rangle / z^2$,
which is twice as large as what we have here.

The reconciliation with Lynden-Bell (1992) is that the quantity he considers
is the {\em mode\/} of $P(\ln(d) | v)$ the distribution of log-distances given an observed recession velocity $v$
and assuming a Gaussian scatter in $v$. That is the most {\em probable\/} log-distance.
But what we are interested in here is the {\em mean\/} of the log-distance.  The $\ln d$ 
probability distribution, under these conditions, is asymmetric, and
the shift of the mean is half the shift of the mode.
Using the shift of the mode, we would argue, overestimates the `homogeneous Malmquist bias'
caused by small scale velocity dispersion by a factor two.

Regarding the analogy with Lynden-Bell et al.\ 1988, what they assumed was a model
for FP distance errors in which the probability distribution for the estimated
log-distance $l_e$ given a true log-distance $l = \ln d$ was a Gaussian:
\begin{equation}
P(l_e | l) = (2 \pi \Delta^2)^{-1/2} \exp(-(l_e - l)^2 / (2 \Delta^2)).
\label{eq:DLB++88model}
\end{equation}
In the present context redshift $z$ plays the role of estimated distance, 
with $v$ the distance error.
But the model (\ref{eq:DLB++88model}) differs from that assumed above 
(with a Gaussian distribution
for velocity errors) in two respects: First, this distribution implies
an asymmetric distribution for the peculiar velocity, with a non-zero mean and
asymmetric tails.  Second, the fractional distance error is independent of
distance, so in this model the absolute error grows with distance.  
This is appropriate for TF or FP distances, but not for errors produced
by random motions.  As we show in the appendix, the former does not,
by itself, resolve the inconsistency;
if one uses the moments of $v$ implied by this distribution 
in (\ref{eq:delta_DM}) this gives $\delta \mu = (5 / \ln 10) \Delta^2$, which
does not agree with (\ref{eq:delta_mu_incoherent}) nor, for that matter,
is it in accord with $\delta \ln d = 3 \Delta^2$.  
The full resolution, again demonstrated in the appendix, is that one needs to modify
the above argument to treat the case of distance independent {\em fractional\/} distance errors, 
and the bias is then given by (\ref{eq:generalisedHMB})
which is very similar to (\ref{eq:delta_DM}) but which has the 
numerical factor $\gamma + 3/2$ replaced by 7/2.
Using the first and second velocity moments implied by (\ref{eq:DLB++88model}) 
in (\ref{eq:generalisedHMB})
gives $\langle d \mu | z \rangle = (15 / \ln(10)) \langle v^2 \rangle / z^2$  in
accord with the usual formula $\delta \ln d = 3 \langle v^2 \rangle / z^2$.  But this
is not correct for distance errors from velocities, which is what we are considering 
here, where it is the {\em absolute\/}
rather than fractional distance error that is independent of distance,
and where the velocity distribution is symmetric.

The above argument is idealised in that it assumes both the density of galaxies and
the velocity distribution function to be independent of position.  Regarding the homogeneous
Malmquist bias the effect of relaxing this is that the expectation of the sky-averaged $\overline{\delta \mu}$
involves the galaxy weighted velocity variance.  For large-scale density perturbations there
is also an inhomogeneous Malmquist bias term (whose expectation vanishes), just as found by Lynden-Bell (1992).
In this regard, we note that the variable $\Phi^{3/2 + d \ln \phi / d \ln z}$ is only 
unbiased with respect to the homogeneous Malmquist bias and is still affected
by the inhomogeneous Malmquist bias.

\section{Summary}

We have shown in \S\ref{sec:coherentflows} that the relatively large low-redshift
perturbations to the mean distances, flux densities or distance moduli obtained from relativistic
second order perturbation theory can be understood as a purely classical kinematic
and statistical Malmquist-like effect and are not, in any essential way, a manifestation
of non-linear dynamics.  While gravity is involved in generating
peculiar velocities, precisely the same bias would be found if one were
observing test particles with non-gravitationally generated motions.  
The relativistic treatment may contain other effects
that are essentially gravitational in nature, but as they are apparently 
extremely small they are of limited interest.

In \S\ref{subsec:galaxy_average} we generalised the
analysis to obtain the bias when, as in reality, the distance
is averaged over sources such as galaxies or supernovae that are subject to selection bias.

Our analysis provides formulae that could, in principle, be used to 
correct for biases in distance, and hence in the `local' value of $H_0$,
from large-scale or small-scale motions.  For the former, our results
properly account for covariance and suppression
of the effect of super-survey modes that is
missing from the relativistic perturbation theory papers.  
But we emphasise that the effects on $H_0$ at least
are very small and much smaller than the fluctuations 
in measurements of $H_0$ that arise in linear theory.

We have shown in \S\ref{sec:incoherentflows} that small-scale 
incoherent velocities have essentially the
same effect.  They do not cause a perturbation to log-distance 
$\delta \ln d = 3 \sigma_v^2 / c^2 z^2$ as has
previously been found, and as would seem reasonable by analogy with the 
commonly used formula for homogeneous
Malmquist bias.  The effect is a factor two smaller.  
The reason that the standard formula
is not valid for bias from velocity dispersion is in part
because the model implies an unrealistic
distribution of velocities and in part because it assumes
that the distance errors scale linearly with distance whereas
errors from motions are distance independent.

We showed that the average of $\Phi^{3/2 + d \ln \phi / d \ln r}$ does
not suffer velocity dispersion induced homogeneous Malmquist bias.

We provide in appendix \ref{sec:generalisedHMB}, a slightly generalised formula
for the homogeneous Malmquist bias produced by errors in estimated luminosity distance -- as from e.g.\ Tully-Fisher 
or fundamental plane techniques -- when using the `forward' method.  
This result is only valid for $\Delta^2 \ll 1$, but makes no assumption
about the form for the distribution function for the distance errors.

\section{Acknowledgements}

The authors thank Ruth Durrer, Dominik Strauss, Giovanni Marozzi and Ido Ben-Dayan for
helpful discussions.
MH acknowledges support of NSERC.
NK gratefully acknowledges the support and
stimulation of the Cifar Cosmology and Gravitation program.

\appendix

\section{Malmquist Bias from Luminosity Distance Errors}
\label{sec:generalisedHMB}

We now obtain the analogue of (\ref{eq:delta_DM}) for the situation where
distances are estimated from the source flux density, rather than redshift, 
and the distance error scales linearly with distance
as is appropriate, to a first approximation, for TF or FP distances or for supernovae.
Thus we assume sources with real distances $d$ and estimated distances $z$ and
distance error $z - d = v$.  I.e.\ we use the same notation as
before, but with a different interpretation as the the cause of the errors,
and to obtain a distance independent distribution for fractional errors we take
$P_v(v|d) dv = f(v/d) dv/d$ where $f(y)$ is some normalised bell-shaped 
function: $\int dy \; f(y) = 1$. 

If we assume the fractional distance errors
are small $v \ll d$, we have
\begin{equation}
\begin{split}
& P_v(v|d) dv = \frac{z}{z-v} f\left(\frac{v}{z-v}\right) \frac{dv}{z} \\
& \quad = \left(1 + \frac{v}{z} + \ldots\right) 
\left(f\left(\frac{v}{z}\right) + \frac{v^2}{z^2} f'\left(\frac{v}{z}\right) + \ldots\right) \frac{dv}{z}
\end{split}
\label{eq:TF+FP_Pv}
\end{equation} 
where $f'(y) = df/dy$.

Our goal is to compute $\langle d | z \rangle$ from the conditional distribution of distance
$P(d | z) \propto P(z|d) P(d)$.  Previously we used $P(d) = d^2 \phi(d)$, but here
the estimated distance is not the redshift, it is the inverse square root of the flux density, so a magnitude
limit imposes a selection that is a function of the estimated distance ($z$ in our notation).  
The upshot, as explained by Strauss \& Willick (1995), 
is that the selection function drops out when we compute $\langle d | z \rangle$
or, equivalently, the bias is the same as obtained without any selection.

Using (\ref{eq:TF+FP_Pv}) in (\ref{eq:delta_DM_int}) we find two extra terms that produce 
significant contributions to the numerator (when multiplied by $-v/z$ and integrated):
\begin{equation}
- z^2 \int \frac{dv}{z} \left(\frac{v^2}{z^2} f\left(\frac{v}{z} \right) + \frac{v^3}{z^3} f'\left(\frac{v}{z}\right) \right)
= 2 \langle v^2 \rangle
\end{equation}
where we have integrated by parts and assumed that $f$ falls to zero for large argument sufficiently fast that
the boundary term is negligible. 

As in (\ref{eq:delta_DM}) one finds, in additional to a term proportional to the variance
in the distance error, the mean distance error $-\langle v \rangle$.  This is not necessarily
zero -- it is not zero, for instance, if the distance estimator is obtained by minimising
residuals in magnitude -- but it is reasonable to assume that 
the strength of any bias in $\langle v \rangle$ is, to order 
of magnitude, at most proportional to the variance $\langle v^2 \rangle$. 
Keeping terms up to quadratic order in the distance error $v$ and ignoring the sub-dominant 
terms in the denominator in (\ref{eq:delta_DM_int}) yields
the general result valid up to linear order in the fractional distance variance $\Delta^2 = \langle v^2 \rangle / d^2$
for the homogeneous Malmquist bias
\begin{equation}
\langle d \mu | z \rangle = \alpha ( - \langle v\rangle / z + 7/2 \langle v^2\rangle / z^2)
\label{eq:generalisedHMB}
\end{equation}
where $v$ is minus the distance error and $z$ can be taken to be either the estimated distance $z$ or the
real distance $d = z-v$.   One can also obtain the perturbation to any other variables.  The perturbation
to the distance, for instance, is
\begin{equation}
\langle {\overline {\delta_d}} \rangle = - \langle v\rangle / z + 4 \langle v^2\rangle / z^2.
\label{eq:generalisedHMBdeltad}
\end{equation}

As a check, we can apply (\ref{eq:generalisedHMB}) and (\ref{eq:generalisedHMBdeltad}) to the log-normal model of Lynden-Bell et al.\ 1988 
that is known to give exactly $\delta \ln d = 3 \Delta^2$ (or $\langle d \mu | z \rangle = (15 / \ln 10) \Delta^2$).
In this model the probability distribution for the estimated log-distance $l_e$ given a true log-distance
$l = \ln d$ is a Gaussian:
\begin{equation}
P(l_e | l) = (2 \pi \Delta^2)^{-1/2} \exp(-(l_e - l)^2 / (2 \Delta^2)).
\end{equation}
With $d = e^l$ and $z = e^{l_e}$ the moments of
the estimated distance distribution are $\langle z^n \rangle = d^n \exp(n^2 \Delta^2 / 2)$.
The first moment is $\langle z \rangle = d (1 + \Delta^2 / 2 + \ldots)$, 
so the mean of the distance error is $\langle v \rangle = d \Delta^2 / 2 + \ldots$, which is non-zero,
and the second moment is $\langle z^2 \rangle = d^2 (1 + 2 \Delta^2 + \ldots)$ 
so the distance error variance is $\langle v^2 \rangle = d^2 \Delta^2 + \ldots$, 
where the notation $\ldots$ indicates quantities
that are of higher order in the assumed small logarithmic variance $\Delta^2$.
Using these in (\ref{eq:generalisedHMB}) gives 
$\langle d \mu | z \rangle = (15 / \ln 10) \Delta^2$
in agreement with Lynden-Bell et al.\ 1988, while (\ref{eq:generalisedHMBdeltad})
gives ${\overline {\delta_d}} = (7/2) \Delta^2$, in accord with equation 185 of Strauss \& Willick (1995).

Equations (\ref{eq:generalisedHMB}) and (\ref{eq:generalisedHMBdeltad}) provide a
generalisation of the standard results in that they do not assume a perfectly log-normal
distribution, though they are limited to the regime where the fractional variance
$\Delta^2 \ll 1$.  They apply only to the `forward' method where one averages the
peculiar velocity of objects as a function of estimated distance.  The more popular
`inverse' methods do not suffer this bias.  Instead they have the much smaller
residual bias from random motions  causing scatter in distance that is the
focus of the main part of this paper. 

\end{document}